\documentstyle{lamuphys}
\makeatletter
\let\chapter\hid@chapter
\makeatother

\def\micron{\hbox{$\mu$m}}
\def\laeq{\lower.5ex\hbox{{$\:\scriptstyle\buildrel < \over \sim\:$}}}

\input{psfig.tex.mine}

\begin{document}
\pagenumbering{arabic}
\title{Introduction to Interferometry}

\author{Timothy R.\,Bedding}

\institute{School of Physics, University of Sydney 2006, Australia}

\maketitle

\begin{abstract}
This tutorial gives a general introduction to optical and infrared
interferometry.  The observables measured by an interferometer are fringe
visibility and closure phase, which can then be fitted by a model or used
to reconstruct an image.  In order to measure fringes, the path lengths
from the source to the detector via the different individual apertures must
be equalized.  This requires some form of fringe tracking, which constrains
the type of observations that are feasible and has important implications
for the complexity and size of the resulting image.
\end{abstract}

\section{Basic principles}

The purpose of these tutorials is address two questions: ``Can I use VLTI
to observe my favourite object?'' and ``What will it tell me?''  Technical
details of interferometry will only be discussed in so far as they affect
the scientific capabilities.  For more details, the reader is referred to
reviews by \cite{Woo82}, \cite{Rod88} and \cite{S+C92}.

\subsection{Angular resolution}	\label{sec.resolution}

We begin with the fact that big telescopes produce sharper images.
Neglecting atmospheric effects, the angular resolution (in radians) of a
telescope of diameter $D$ is given by
\begin{equation}
 \theta = \lambda / D.	\label{eq.theta}
\end{equation}
This omits a factor of 1.22 appropriate to a circular aperture, which we
dismiss as a technical detail, and neglects any image degradation from
imperfections in the optics.  The following table shows typical angular
resolutions in milliarcseconds (mas) for three wavelength regimes:

\bigskip
\begin{tabular}{l|ccc}
	& Visible	& Near IR	& Mid-IR	\\
	& ~~0.4--1\,\micron~~ & ~~1--5\,\micron~~ & ~~10--20\,\micron~~ \\ \hline
D=8\,m	& 18\,mas	& 77\,mas	& 400\,mas	\\
D=100\,m~& 1.4\,mas	& 6\,mas	& 30\,mas	
\end{tabular}
\bigskip

\begin{figure}
\centerline{
\psfig{figure=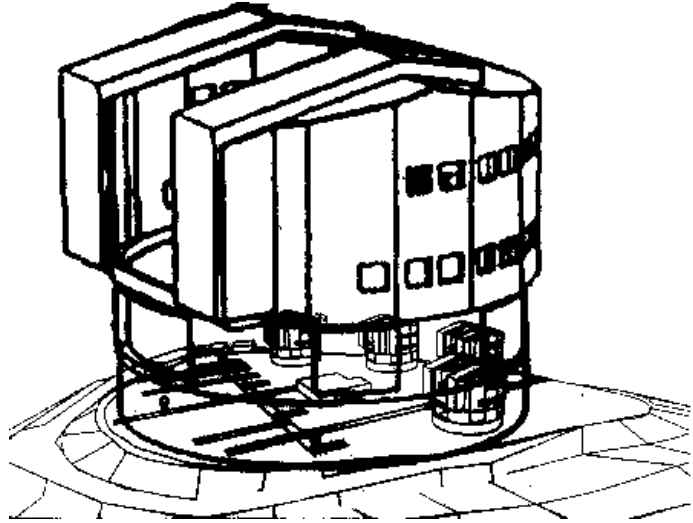,width=\the\hsize,scale=75}}

\caption[]{\label{fig.paranal} An expensive way to achieve high angular
resolution from Paranal.}
\end{figure}

Interferometry allows us to achieve high angular resolution without
building a telescope of enormous size (Figure~\ref{fig.paranal}).  Instead,
we synthesize a large aperture by combining light beams from several small
telescopes.  When we observe a distant object in this way, we see
interference fringes.  These fringes arise because of the wave nature of
light and they contain information about the object being observed.

\subsection{Fringe visibility}

The visibility of fringes is a number between zero and one which measures
the fringe contrast (see Figure~\ref{fig.fringes}).  It is defined as $V =
(I_{\rm max} - I_{\rm min})/(I_{\rm max} + I_{\rm min})$.  If the dark
regions in the fringe pattern go down to zero, the fringes have a
visibility of one and we say the object is unresolved.  If $V=0$ there are
no fringes and the object is completely resolved.

It is important to remember that interferometers record fringes, not images
(unlike adaptive optics systems, which {\em do\/} record images).  So the
question ``Can I use VLTI to observe my favourite object?''  becomes ``Can
I measure fringes from my object?''  If the object is very large and
diffuse, the answer may be no.

\subsection{Interpreting fringe visibility}

In practice, fringes from an interferometer will never have perfect
contrast ($V=1$), even for a point source, because of atmospheric and
instrumental effects.  One must therefore accompany each measurement of the
target by a similar measurement of an unresolved calibrator source.  The
`true' visibility of the object is then the observed visibility of the
object divided by the observed visibility of the calibrator.

\begin{figure}
\centerline{
\psfig{figure=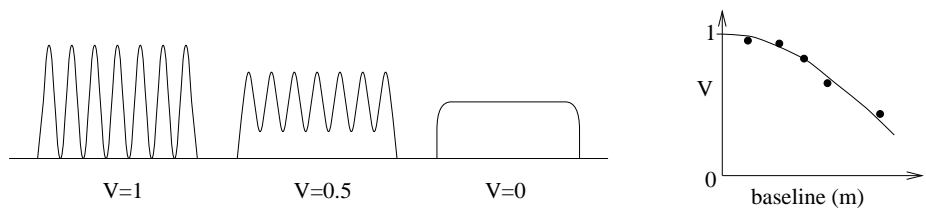,width=\the\hsize,scale=100}}

\caption[]{\label{fig.fringes} Left: examples of fringes with visibilities
of 1, 0.5 and 0.  Right: visibility as a function of baseline for a
resolved star.  }
\end{figure}

As an example, consider a two-telescope interferometer which is measuring
the angular size of a star.  The procedure is to measure $V$ at different
baselines, one at a time, by moving the telescopes.  From the sequence of
measurements we find that the fringe visibility goes down as the baseline
increases (see Figure~\ref{fig.fringes}).  The star is becoming more
resolved on longer baselines.  Fitting a curve to the measured visibilities
allows you to determine the angular diameter of the star.

This visibility curve has a special meaning: it is the Fourier transform of
the object's brightness distribution.  Thus, to make an {\em image\/} of
the sky we need to reverse this process, which means performing an inverse
Fourier transform of the measured visibilities.  To do this accurately for
anything but the most simple object requires that you sample the visibility
at many different baselines.  We talk about needing good coverage of the
$(u,v)$ plane, a phrase borrowed from radio astronomers.  Having more
telescopes in your interferometer saves time, since it allows you to
measure visibilities on several baselines simultaneously (the number of
baselines from an array of $M$ telescopes is $M(M-1)/2$).  Allowing the sky
to rotate during the night also improves $(u,v)$ coverage, and this
technique is called earth rotation synthesis.



\subsection{Closure phase}

As stated above, the fringe visibility measured using a particular baseline
tells you one component of the object's Fourier transform.  However, a
Fourier transform is a complex quantity, having both an amplitude and
phase.  While the amplitude is given by the fringe visibility, the phase is
given by the position of the fringes.

Unfortunately, the fringes from an interferometer continually move
backwards and forwards because of atmospheric turbulence, which changes the
path length in front of each aperture.  Therefore, it seems that the
position of the fringes tells you nothing about the object.  However,
consider the case of three telescopes.  We can observe three sets of
fringes simultaneously, all moving, but they don't move independently.
Their movement is due to the atmosphere, but their {\em relative\/}
positions contain information about the object.  Thus, with three
telescopes you can measure an additional piece of information, called the
closure phase, which contains some (but not all) of the phase information.
For example, the closure phase tells you about the asymmetry of the object
(but not its absolute position on the sky).  An array of more than three
telescopes yields a closure phase for each triangle, although not all these
measurements are independent.  Combining these closure phases with the
visibility measurements allows you to reconstruct an image of the object.

\section{Interferometry in practice}


\subsection{Atmospheric effects}

An interferometer measures fringes to obtain visibilities and closure
phases.  Why is this so difficult?  A plane wavefront from a distant star
travels through the atmosphere and is distorted by fluctuations in
refractive index.  The wavefront that reaches our ground-based telescope is
no longer flat.  The angle between different parts of the wavefront is
about one arcsecond, which is why the image seen by a single large
telescope is a fuzzy blob about one arcsecond in size.  A useful analogy is
to represent the distorted wavefront by a crumpled piece of paper.  An
adaptive optics system tries to flatten the paper using a large number of
fingers (actuators).  On the other hand, an interferometer takes small
pieces of the wavefront and combines them.

Note that the shape of the wavefront is approximately the same at all
wavelengths: it has the same physical size in microns (because the
refractive index of air varies only weakly with wavelength).  That is why
the seeing in arcseconds is approximately the same at all wavelengths.  Why
then is interferometry easier in the infrared?  The important thing is the
size of the wavefront measured not in microns, but in wavelengths.  As the
wavelength increases, the distortions become relatively less important.

We define the atmospheric coherence length $r_0$ to be the size of a `flat'
patch on the wavefront, where this is decided relative to the wavelength.
More precisely, $r_0$ is defined so that the rms variation in the wavefront
across a patch of that diameter is $\frac{\lambda}{2\pi}$ (one radian of
phase).  In typical seeing $r_0$ is 10--20\,cm at visible wavelengths and
becomes larger at longer wavelengths.  The increase is not exactly linear,
with $r_0$ depending on wavelength as $\lambda^{6/5}$.

What about time variation?  The shape of the wavefront changes with time,
but to good approximation the dominant effect is the wind blowing the whole
pattern past the telescope.  This model is known as frozen turbulence.  We
can then define an atmospheric coherence time $\tau_0$, which is roughly the
time taken for one $r_0$-sized patch to move past.  In the visible this is
a few tens of milliseconds, again increasing with wavelength as
$\lambda^{6/5}$.

\subsection{How does an interferometer work?}

\begin{figure}
\centerline{
\psfig{figure=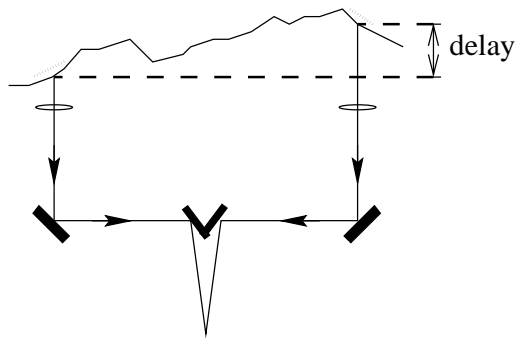,width=\the\hsize,scale=70}}
\caption[]{\label{fig.wavefront} The wavefront above a two-element
interferometer, showing the corrections required for tip-tilt and delay }
\end{figure}

An interferometer combines two or more separate parts of the wavefront to
produce fringes.  This requires two corrections to the wavefront, as shown
in Figure~\ref{fig.wavefront}.  Firstly, although the wavefront across each
individual telescope is approximately flat, it will have a tilt, so we
require tip-tilt correction to make the beams overlap.

The second correction is to make the path lengths equal, without which no
fringes will be visible.  The crude part of this correction is to allow
observations of objects all over the sky, not just those which are directly
overhead.  This is achieved by delaying one beam (by bouncing it back and
forth along a tunnel), which allows you to steer the interferometer.  Much
harder is to make the small correction for the atmosphere, a few tens of
microns, which varies rapidly and randomly.  This explains the need for
fringe tracking.

\subsection{Fringe tracking}

To measure fringes, we must maintain equal optical paths for a reasonable
period (minutes at a time).  There are three ways to achieve this:
\begin{enumerate}

 \item track fringes on the scientific target (which requires the target to
have a bright compact component)

 \item track fringes on a nearby reference star (note that a laser beacon
is {\em not\/} compact enough)

 \item blind tracking (which requires a very stable and well-calibrated
system; this may be possible in the mid-IR)

\end{enumerate}

In addition, there are two {\em levels\/} of fringe tracking.  The more
difficult to achieve is `cophasing,' which means maintaining equal optical
paths to within a fraction of a wavelength.  The fringes will then be
steady, allowing integration for many times~$\tau_0$.  This is clearly
useful in the case of a bright reference star and a faint and/or resolved
target: fringes can be tracked to high accuracy on the reference star,
allowing long integrations on the target.

Less difficult is to track to within the coherence length of the light
($\lambda^2/\Delta\lambda$), so that the fringes are still detectable but
are continually moving.  In this case, called `coherent' tracking, you must
measure the fringes in many short exposures ($\tau_0$).

\begin{figure}
\centerline{
\psfig{figure=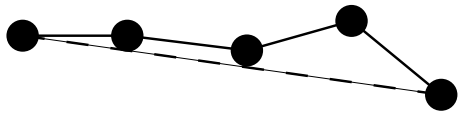,width=\the\hsize,scale=50}}

\caption[]{\label{fig.bootstrap} Fringe tracking by bootstrapping.  The
dots are the individual telescopes.  On the short baselines (solid lines)
the target is only partially resolved and fringe tracking should therefore
be easy.  By tracking on each of these short segments, the fringes on the
longest baseline (dashed line) will be stable and can be measured, despite
their low visibility.  }
\end{figure}

There are some tricks which help with fringe tracking.  One is
bootstrapping, which involves tracking fringes on a series of short
baselines (see Figure~\ref{fig.bootstrap}).  A possible problem is the
propagation of errors, perhaps resulting in an unacceptably large error on
the long baselines.  Of course, it also requires at least four telescopes
to make the gain worthwhile and, in terms of $(u,v)$ coverage, these must
be deployed rather wastefully.

Another trick is to track at one wavelength and measure fringe visibilities
at another.  For example, at longer wavelengths the object should be less
well resolved (Equation~\ref{eq.theta}), so you can track at a long
wavelength and observe at a shorter one.  For some objects (e.g., Be stars)
you can track on a compact continuum source and observe extended structure
in an emission line.

Nevertheless the ability to track fringes is a difficult requirement and
will usually set the limiting magnitude of the interferometer.

\section{What can we expect?}

\subsection{Image complexity}

We should not expect to produce complex images like those from the VLA and
similar radio interferometers.  The best analogy between radio and optical
interferometry is VLBI, which use six or seven apertures over
Intercontinental baselines.  An array of 27 radio telescopes gives
$27\times26/2 = 351$ simultaneous baselines, allowing very efficient
coverage of the $(u,v)$ plane.  This works well in the radio because you
can amplify the signal from each telescope before combining them.
Unfortunately, in the visible and near infrared this cannot be done without
introducing noise that swamps the signal.  Every extra telescope adds
noise, and there comes a point where it is detrimental to add more
telescopes.  For fuller discussion of these points see \cite{Nit94} and
\cite{Pra94}.

Of course, extra telescopes could be used in a completely separate array,
by combining their light on a second detector.  But then the number of
baselines being measured simultaneously will be much less than $M(M-1)/2$.
For this reason, no optical/infrared interferometer is being built with
more than six elements.  It is therefore unlikely that image complexity
will match that of the VLA\@.  Roughly speaking, the number of non-zero
resolution elements in the final image equals the number of independent
measurements in the $(u,v)$ plane.  For VLTI this will be a few tens to a
few hundreds.  Finally, for on-source fringe tracking there is an extra
constraint on image complexity, namely that the object have a compact
component.

\subsection{Field of view} 

How big will the final image be?  One restriction comes directly from the
preceding discussion on image complexity.  The {\em number\/} of non-zero
pixels is limited by $(u,v)$ coverage while their {\em size\/} is set by
the resolution of the interferometer.  Since this resolution is typically
reckoned in milliarcseconds, the images of a complex object (one with many
non-zero pixels) cannot be very large.

There is another restriction on the field of view.  Equal path lengths,
necessary to detect fringes in the first place, must be maintained over the
whole field.  Fringe tracking will (hopefully) ensure that paths are equal
at the centre of the field.  But looking off-axis by an angle $\theta$
introduces an extra path length of $D\theta$, where $D$ is the length of
the baseline.  This extra path difference will only be unimportant if it is
small compared to the coherence length of the light
($\lambda^2/\Delta\lambda$).  Thus, the limit on the field of view is
\[
 \theta \laeq \frac{\lambda}{D} \frac{\lambda}{\Delta\lambda}.
\]
Writing it in this way, we see that the field of view is set by the product
of the angular resolution and the spectral resolution.  This limits the
size of the region over which fringes can be detected.  For example, if the
field width is to be 100 resolution elements, which would be about
0.6\,arcsec for VLTI in the near IR, the spectral resolution must be better
than 100.

In principle this limit can be overcome by building a Fizeau interferometer
(also called homothetic mapping in ESO publications), which requires very
complicated extra optics to ensure path equality over the whole field.
Another approach is to build a separate delay compensator for each part of
the field.  This is done for a dual-feed system, in which one (small) field
is centred on the target and the other on the reference star.  In this
case, the maximum separation between object and reference is set by
isoplanatism.

\subsection{Dynamic range} 

This is the ratio of the brightest part of the image to the faintest part.
Images with 100:1 are already possible with optical interferometers, while
1000:1 will be difficult.  This must be kept in mind when deciding what to
expect from VLTI observations of your favourite object.

\section{Conclusion} 

Interferometers give improvement in angular resolution of several orders of
magnitude over current ground- and space-based telescopes, but at a price.
They cannot produce beautiful images of complex objects like those from HST
or even the VLA\@.  Despite the difficulties and limitations, the enormous
potential gains in resolution have spurred more than a dozen major
interferometry projects around the world.  The latest information on these
can be found at \verb"http://huey.jpl.nasa.gov/~shaklan/olbin/" (Optical
Long Baseline Interferometry News).

\subsection*{Acknowledgements} 

I have benefited greatly from discussions with many colleagues over the
years, and would particularly like to acknowledge Gordon Robertson, Ralph
Marson and Oskar von der L\"uhe.  I also thank Gordon Robertson for
comments on this paper.  For financial assistance I am grateful to the
University of Sydney Research Grants Scheme and the Australian Research
Council.

%

%
%

\subsection*{DISCUSSION}

\small
\tt
\parindent 0em
\parskip 1ex
\raggedright

Theo ten Brummelaar: I think you glossed over the calibration problem just
a little bit too much.  At the longest baselines, finding unresolved
objects will be difficult.

Tim Bedding: The object you use as a calibrator doesn't have to be
unresolved, it just has to be known and preferably smaller than the target
object.  But you are right, at the longest baselines almost everything you
look at will be resolved.  You have to bootstrap, calibrating on partially
resolved stars.  This is more of problem in the visible than in the near
and mid infrared, where the resolution is less.

Andreas Quirrenbach: Another comment on calibration.  You are quite right
that this can be a big problem and not so much because the calibrator is
resolved but because of variations in time and over the sky.  If you want
to have an image with a dynamic range of 100:1 or 1000:1, you also require
that precision in calibration of visibility measurements.  You have to
measure fringe contrast to a precision of 1\% or 0.1\% through the changing
atmosphere.  It's basically the same problem as calibrating the
point-spread function in adaptive optics observations.  The changes on the
timescale of ten minutes are very worrisome and you have to change between
your target object and the calibrator as fast as you can.

Theo ten Brummelaar: If you are cophasing three telescopes, aren't you
servoing out one of the very things you are trying to measure?

Oskar von der L\"uhe: If you have three telescopes, it is sufficient to
stabilize fringes between two pairs.  The phase on the third baseline will
be determined by the object.  So, no, you are not destroying a good
observable when you fringe track.

Martin Ward: How good could the astrometry be in principle?

Mike Shao: Tens of microarcseconds

George Miley: You mentioned that adding baselines wouldn't improve things.
But of course the dynamic range is limited by the number of $(u,v)$ points,
which is limited by the number of telescopes.  Adding baselines helps
dynamic range, at least for bright objects, even though you do add noise.

Tim Bedding: As far as image making is concerned I agree, the more
telescopes the better.  I'm talking here about signal detection: the need
to detect fringes and track them.  If you build an optical array with too
many telescopes, you won't be able to measure fringes on any baseline,
because you are adding photons from all of those telescopes at once.  [The
observed fringe visibility on a given baseline will decrease because the
fringes are seen against the background light from all the other
telescopes.]

Chris Haniff: If you're doing regular astronomy, most astronomers agree
that if you have bigger telescopes you get better data.  If you are
building an interferometer and someone gives you extra money, what would
you do?  Should you build more telescopes or bigger telescopes?

Tim Bedding: Making the telescopes bigger only helps up to a point.  Once
the telescope diameter reaches a few times $r_0$, you don't gain nearly as
quickly.  For example, the VLTI Auxiliary Telescopes (ATs) are designed to
be 1.8\,m because that is a few times $r_0$ in the near infrared.  You
wouldn't try to operate a visible interferometer with 2\,m telescopes.  So
the answer depends on the wavelength.  If you want to operate at
10\,\micron\ then make them 8 metres, since at this wavelength an 8\,m
aperture is almost diffraction limited (Section~\ref{sec.resolution}).  The
wavefront across those Unit Telescopes on Paranal is almost flat in the
mid-IR, which is why it's such a great interferometer.  So if you want to
operate in the mid-IR you would build one or two more UTs.  However, if you
want to operate in the near-IR you want more ATs, five or six of them.

Chris Haniff: But not twenty?

Tim Bedding: No.

Wesley Traub: If you had the money, another possibility is to invest in
making a laser guide star and adaptive optics for each telescope.

Tim Bedding: Yes, it would be great to have adaptive optics for the Unit
Telescopes in the near infrared.  The other thing you could buy is one of
these superconducting tunnel junction detectors for fringe tracking.

Francesco Paresce: Do you think that in five minutes you could tell us
about phase closure?

Tim Bedding: 
The figure shows three telescopes all looking at the same star.  Above each
telescope there is an unknown and continually changing thickness of
atmosphere.  We measure the phase of fringes between each pair of
telescopes.  By phase, I mean the position of the fringes.  The phases are
continually changing and tell you nothing about the object.

\begin{figure}
\centerline{
\psfig{figure=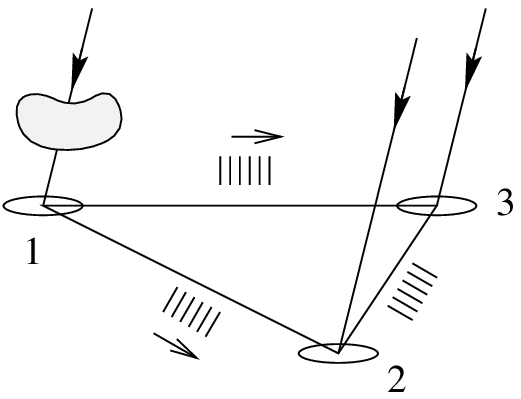,width=\the\hsize,scale=50}}

\end{figure}

Now imagine adding an extra piece of atmosphere in front of Telescope~1.
The fringes on baselines 1-2 and 1-3 will both shift in the direction shown
by the arrows, while the fringes on baseline 2-3 will be unaffected.  What
I measure is the algebraic sum of the three fringe shifts, defining
positive as clockwise round the triangle.  Then the extra atmosphere in
front of Telescope~1 has shifted one set clockwise, one set anticlockwise,
and one set not at all.  The algebraic sum $\phi_{12} + \phi_{23} +
\phi_{31}$ is called the closure phase and, from the preceding argument, it
is completely unaffected by atmospheric fluctuations.  If the atmosphere
disappeared entirely we would get the same number.  This number is telling
you something about the object.

Francesco Paresce:  If you have four telescopes? 

Tim Bedding: From four telescopes you can make four triangles, but only
three of these are independent.  A nice thing is that, unlike with
visibility amplitudes, you don't need to calibrate closure phase.  If you
look at a point source, the closure phase is always zero.  It is much more
resistant to variations in seeing.  And it's also sensitive to object
structure, telling you about asymmetries even below the diffraction limit
of the interferometer.

Jim Beletic: You could add that laser guide stars won't help you phase the
three apertures because the light from the laser spot to each telescope is
coming through a different part of the atmosphere.  You can only use the
laser to flatten the wavefront across each telescope.

Tim Bedding: Yes.  

\end{document}